\documentclass[10pt]{iopart}

\usepackage{graphicx}
\begin{document}


\date{\today}
\title[Perturbation Theory with a Field Cutoff]{A Tractable Example of Perturbation Theory with a Field Cutoff: the Anharmonic Oscillator}

\author{L. Li\dag\ and Y. Meurice\dag\ \ddag \S
\footnote[3]{To
whom correspondence should be addressed (yannick-meurice@uiowa.edu)}
}
\address{\dag\ Department of Physics and Astronomy\\ The University of Iowa\\
Iowa City, IA 52242 USA }
\address{\ddag Kavli Institute for Theoretical Physics, Santa Barbara\\
Santa Barbara, CA 93106 USA }
\address{\S\ Also at the Obermann Center for Advanced Study, University of Iowa}
\date{\today}

\begin{abstract}
For $\lambda \phi^4$ models, the introduction of a large field cutoff
improves significantly the accuracy that can be reached with  
perturbative series but the calculation of the modified coefficients remains 
a challenging problem. We show that this problem can be solved 
numerically, and  analytically in the limits of large and small field cutoffs, for 
the ground state energy of the anharmonic oscillator. 
For the two lowest orders in $\lambda$, the approximate formulas obtained in the large field 
cutoff limit extend unexpectedly far in the low field cutoff region and there is 
a significant overlap with the small field cutoff approximation.
For the higher orders, this is not the case, however the shape of the transition between the small field cutoff regime 
and the large field cutoff regime is approximately order independent.
\end{abstract}
\section{Introduction}

Perturbative methods and Feynman diagrams have played an important role in the development 
of quantum field theory and its applications.  
However, perturbative series usually have a zero radius of convergence\cite{leguillou90}. 
For scalar models with $\lambda \phi^4$ interactions, the 
coefficients of perturbative series grow factorially. For any fixed, 
strictly positive, value of $\lambda$, there exists an order beyond which adding 
higher order terms diminishes the accuracy. 
This feature will restrict our ability to perform high precision tests of the standard model (for instance,  $g-2$ of leptons and the hadronic width of the $Z^0$) during the next decades.

The large order behavior of the
series is dominated by large field configurations which have little effect on low energy observables.
Introducing a large field cutoff \cite{pernice98,convpert} in the path integral formulation of 
scalar field theory, dramatically improves the large order behavior of the perturbative 
series. In two non-trivial examples \cite{convpert},  this procedure
yields series that have finite radii of convergence and tend to values  
that are exponentially close to the exact ones. This also allows us to define the theory for 
negative or complex values of $\lambda$, a subject that has raised a lot of interest 
recently \cite{bender98,gluodyn04}. An important feature of this approach is that 
for a perturbative expansion at a given order in $\lambda$, it 
is possible in some cases to determine 
an optimal field cutoff using the strong coupling expansion \cite{optim03,plaquette}, 
bridging the gap between the two expansions. In other words, the modified perturbative methods allow us to take into account non-perturbative effects. 

Despite these promising features, 
calculating the modified coefficients remains a challenging technical problem. 
While developing a new perturbative method (see e.g. Ref. \cite{buckley92,duncan92} for the $\delta$ expansion), it is customary to demonstrate the advantages of a method 
with simple integrals and the non-trivial, but well-studied\cite{bender69,leguillou90}, case of the anharmonic oscillator. A simple integral has been discussed in Ref. \cite{optim03}.
In this article, we show not only that this program can be completed in the 
case of the anharmonic oscillator, but also that the results show remarkable properties:
\begin{itemize}
\item
For the two lowest orders, the approximate formulas obtained in the large field 
cutoff limit extend unexpectedly far in the low field cutoff region.
\item 
For the higher orders, the transition between the small field cutoff regime 
and the large field cutoff regime can be approximately described in terms of a single 
function.
\end{itemize}

The results are presented in the following way.
In Sec. \ref{sec:num}, we define the model considered and the numerical 
calculation of the 
modified coefficients. In Sec. \ref{sec:rad1}, we discuss the radius of convergence of 
the modified series from a numerical point of view. In Secs. \ref{sec:large} and \ref{sec:small}, we present
the approximate methods at small and large field cutoff. Rigorous bounds 
on the radius of convergence of 
series encountered in the previous sections are given in  Sec. \ref{sec:rad2}.  The question of the interpolation 
between the small and large field cutoff regimes is discussed in Sec. \ref{sec:crossemp}.
The importance of a complete understanding of this question, as well as the extension to field theory are discussed in the conclusions.

\section{The problem and its numerical solution}
\label{sec:num}
In this section, we introduce the anharmonic oscillator with a ``field cutoff'' and we explain 
how to calculate numerically the perturbative series in the anharmonic coupling 
$\lambda$. This method was first used in \cite{convpert} and tested with the known results \cite{bender69} up 
to order 20. It is a perturbative version of the numerical method proposed 
in Ref. \cite{arbacc}.
For convenience, we use quantum mechanical notations 
instead of field theoretical ones 
$
\phi \rightarrow x , \ 
 m \rightarrow \omega $ and the field cutoff will be denoted $x_{nax}$.
We also use units such that $\hbar=1$ and the ``mechanical mass'' $m$ is 1. However, the harmonic angular frequency $\omega$, will sometimes be used as an expansion parameter and will be kept arbitrary in the equations. In these units, $x_{max}\sqrt{\omega}$ and 
$\lambda/\omega^3$ are dimensionless.
The hamiltonian reads
\begin{equation}
H=\frac{p^{2}}{2}+V(x)
\end{equation}
with
\begin{equation}
V(x)=\left\{
\begin{array}{ccc}
\frac{1}{2}\omega ^{2}x^{2}+\lambda x^{4}\quad & {\rm if}& |x| < x_{\max}  \\
\infty \quad & {\rm if} & |x| \geq x_{\max}
\end{array}
\right.
\end{equation}
Our main goal is the calculation of the modified coefficients $E_0^{(k)}(\sqrt{\omega}x_{max})$ of the perturbative series for the 
ground state energy:
\begin{equation}
E_0(x_{max},\omega,\lambda)=\omega \sum_{k=0}^{\infty }E_0^{(k)}(\sqrt{\omega}x_{max})\times (\lambda/\omega^3) ^{k}\ ,
\label{eq:eexp}
\end{equation}

For this purpose, we will solve perturbatively the time independent Schr\"odinger equation 
with the boundary condition $\Psi(x_{max})=0$. 
We proceed as in Ref. \cite{arbacc}. Setting
\begin{equation}
\Psi(x)\propto{\rm e}^{-\int_{x_0}^{x}dy (L(y)/K(y))} \ ,
\label{eq:repa}
\end{equation}
the Schr\"odinger equation reads 
\begin{eqnarray}
\label{eq:basic1}
 L'&+&2(V-E)K+GL=0 \\
\label{eq:basic2} K'&+&L+GK=0
\end{eqnarray}
where $G(x)$ is an arbitrary  function. 
The second of these equations implies that 
\begin{equation}
\Psi(x)\propto K(x){\rm e}^{\int^x dy G(y)}
\end{equation}
It possible to show \cite{arbacc} that if $G$ and $V$ are polynomials:
\begin{itemize}
\item
 Eqs. (\ref{eq:basic1}-\ref{eq:basic2}) can be solved by power series which define entire functions; 
 recursion formulas and initial conditions (different for even and odd 
 solutions) are given in Ref. \cite{arbacc}.
 \item
  the 
zeroes of $\Psi$ are the sames as the zeroes of $K$. 
\end{itemize}
This implies that the energy levels can be obtained by solving 
\begin{equation}
K(x_{max},\omega , \lambda, E)=0
\label{eq:kzero}
\end{equation}
for the variable $E$, and that polynomial approximations can be used for this purpose. 
We now use the perturbative expansion 
\begin{equation}
E=\omega \sum_{k=0}^{\infty }E^{(k)}(\lambda/\omega^3) ^{k}\ .
\label{eq:eexpgen}
\end{equation}
We assume that $\omega>0$. The case $\omega=0$ is simpler and will be discussed in 
Sec.  \ref{sec:small}. By construction, the coefficients  $E^{(k)}$ are dimensionless
and can only depend on $\sqrt{\omega}x_{max}$ when their value is fixed using Eq. 
(\ref{eq:kzero}). In the following, this dependence will sometimes be kept implicit in order to reduce the size of some equations.
We also recall that if we had not chosen the units   
$\hbar=m=1$, the dimensionless quantities used above would read $(\sqrt{\omega m/\hbar}x_{max})$ and $\hbar \lambda/m^2\omega^3$.

We need to set the expansion
\begin{eqnarray}
K(x_{max},\omega , \lambda, E)&=&K^{(0)}(\sqrt{\omega}x_{max},E^{(0)})+\\
&\ &(\lambda/\omega^3)	K^{(1)}(\sqrt{\omega}x_{max},E^{(0)},E^{(1)})+\dots 
\end{eqnarray}
equal to zero order by order in $\lambda$.
We can approximate the $K^{(k)}$ by polynomials. We then start by solving  $K^{(0)}(\sqrt{\omega}x_{max},E^{(0)})=0$ for $E^{(0)}$ using Newton's method.
This correponds to the problem of an harmonic oscillator potential that becomes 
infinite at $x=\pm x_{max}$. The various zeroes of the polynomial 
correspond to the (even or odd depending on how we construct $K$ \cite{arbacc}) spectrum of this model.
The ground state energy is obtained by taking the lowest even solution.
By increasing the degree of the polynomial approximation, we can stabilize the numerical value with great accuracy. 
The independence on $G$ of the exact equations, before we use polynomial approximations, 
can be used to test the numerical accuracy of the polynomial approximations used 
for $K$. In practice a good choice is made by having $K$ of order 1 
for $E$ close, but not fine tuned, to its actual value.
Since the potential is even, it is natural to have $K$ even and $G$ and $L$ odd, 
however we noticed that introducing a small parity breaking in $G$ 
usually improves the numerical stability.
We then solve $K^{(1)}(x_{max},E^{(0)},E^{(1)})=0$ for $E^{(1)}$. Since  $E^{(1)}$ appears only
linearly in the order $\lambda$ expansion of $K$, it is a linear equation for this 
quantity and it can be solved trivially. 
The same reasoning shows that the higher order equations are linear in $E^{(k)}$.

\begin{figure}\begin{center}  
\label{fig:num}               
\includegraphics[width=0.6\textwidth]{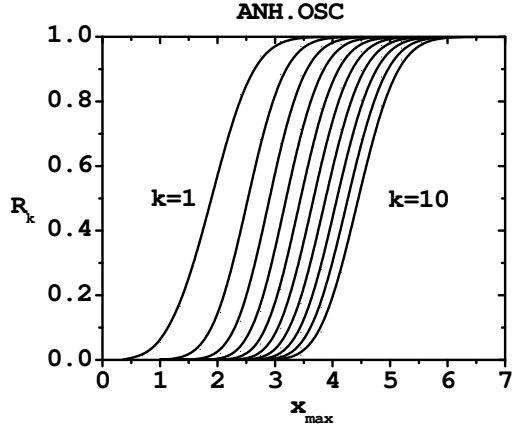}
\caption{The numerical values of $R_{k}(x _{\max
})
$ for $k=1, \dots 10$ and $\omega=1$.\label{fig:an-ratio}} 
\end{center}
\end{figure}

By using this method, we have calculated the first ten 
coefficients for $\omega=1$ and values of $x_{max}$ between $0.5$ and 7.
In order to allow a comparison among the different orders, we define the ratios
\begin{equation}
R_k(\sqrt{\omega}x_{max})\equiv E_0^{(k)}(\sqrt{\omega} x_{max})/E_0^{(k)}(\infty)\ ,
\end{equation}
which all tend to 1 in the 
$x_{max}\rightarrow\infty$ limit and to 0 in the $x_{max}\rightarrow 0$ limit. The numerical values of these ratios are shown in 
Fig. \ref{fig:an-ratio}. A striking feature is that the curves for the various orders have approximately the same shape and that we could approximately superpose them by appropriate horizontal translations. 
Before studying these curves in the large and small $x_{max}$ limits, we will discuss 
numerical estimates for the radius of convergence of the modified series, that can be extracted from this data.

\section{Numerical evidence for a finite radius of convergence}
\label{sec:rad1}

If we calculate the integral of a function defined by a series which is 
uniformly convergent over the range of integration, it is legitimate to interchange the sum and the integral. On general grounds, one would thus expect that if we calculate the partition function 
of a properly regularized theory (say on a finite lattice) 
with a field cut, the perturbative expansion becomes convergent for arbitrary  complex values of the coupling. The ground state energy of the anharmonic 
can be obtained from the logarithm of the partition function and consequently, 
we would expect that its perturbative expansion will have a finite radius of convergence. This radius will depend on the position of the complex zeroes 
of the partition function. 
\begin{figure}[t]
\begin{center}
\includegraphics[width=0.6\textwidth]{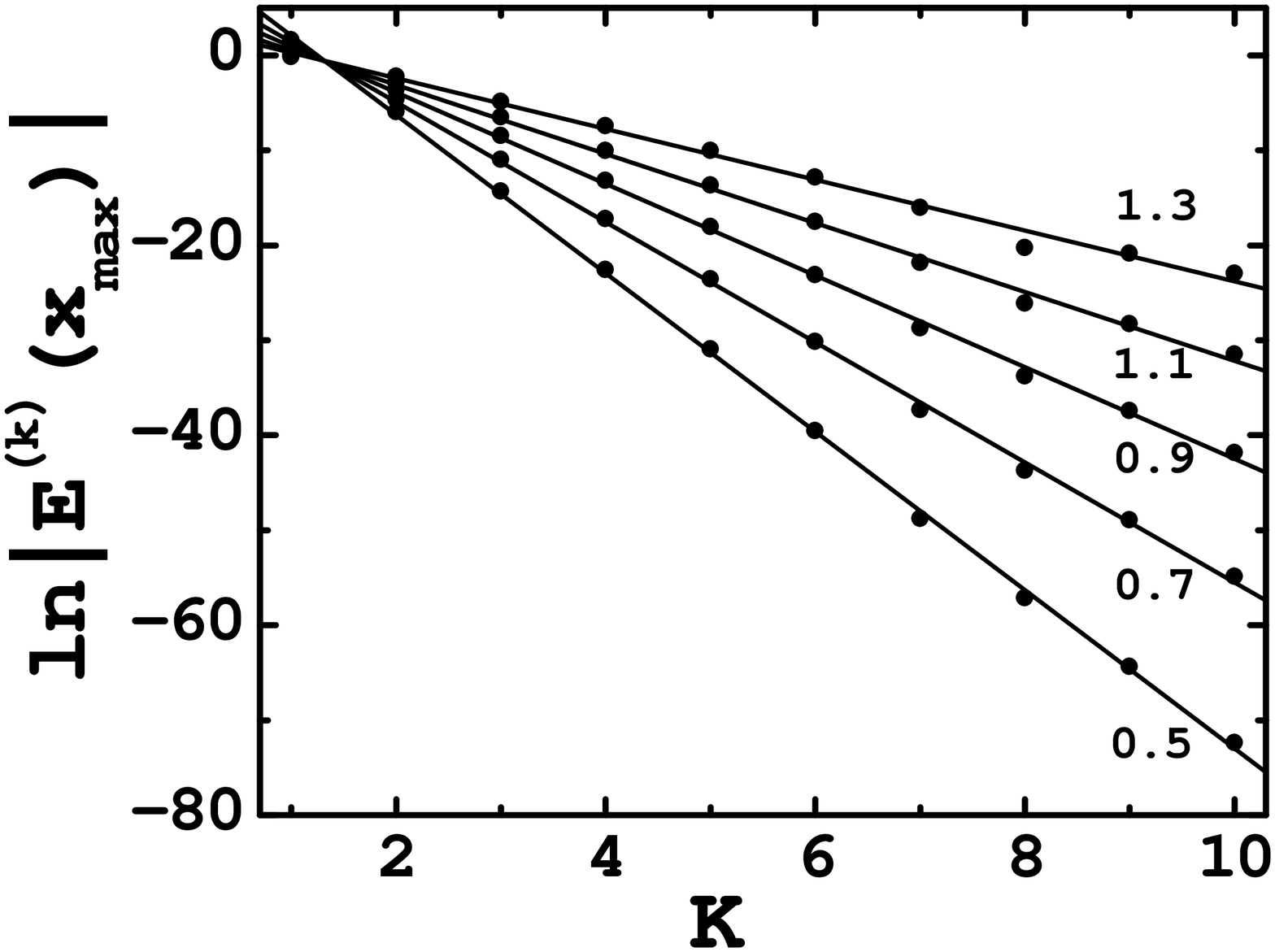}
\caption{ln$|E^{(k)}_0(x_{max})|$ versus $k$ for $x_{max}$=0.5, 0.7, ...1.3 (from bottom to top); $\omega$ was set to 1. \label{fig:logek}}
\includegraphics[width=0.6\textwidth]{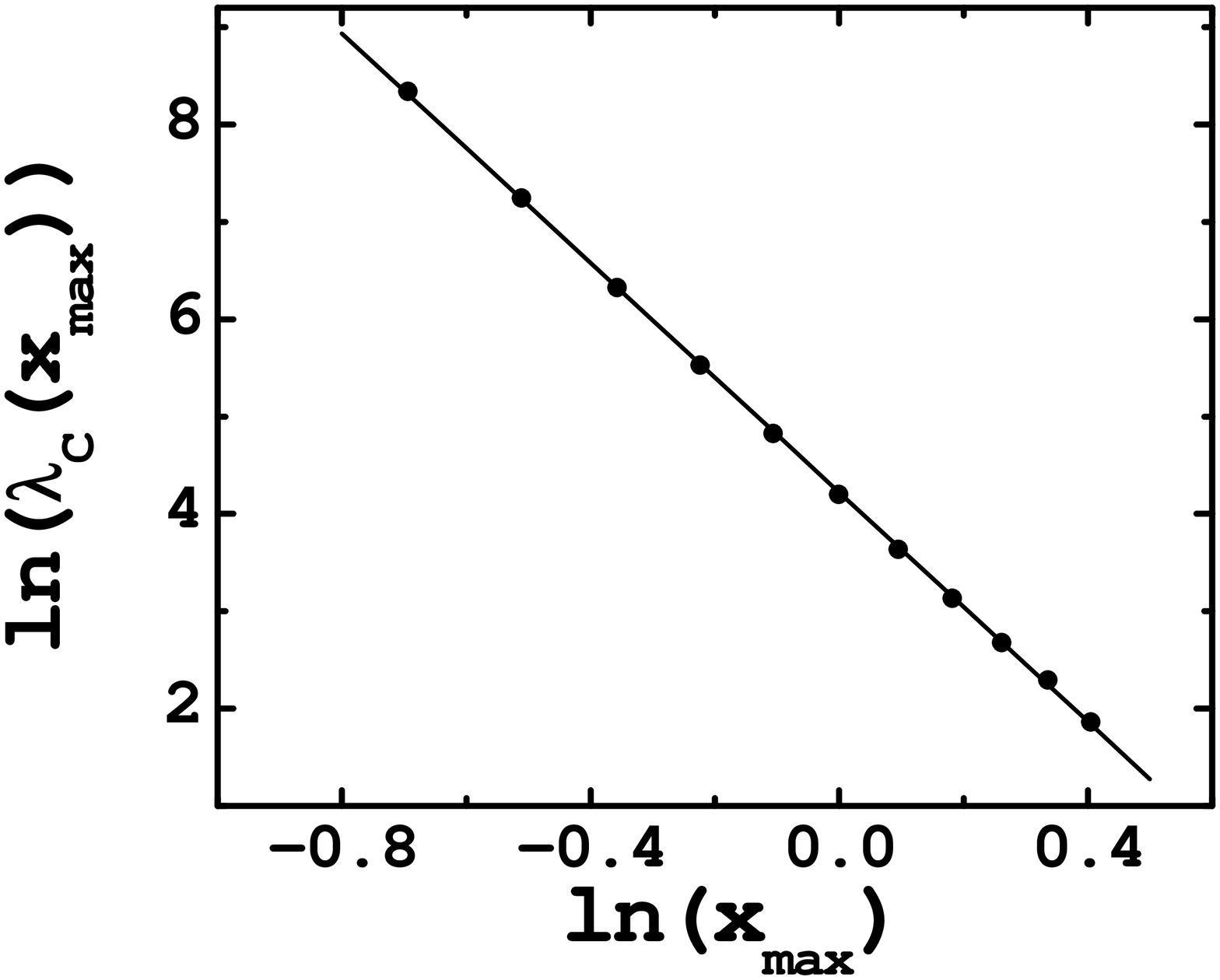}
\caption{ln$\lambda_c(x_{max})$ versus ln$x_{max}$ for $\omega=1$. The line is $4.22- 5.89 x$. \label{fig:lambdac}}
\end{center}
\end{figure}

It is sometimes possible to estimate the radius of convergence by considering the 
empirical asymptotic behavior of the perturbative series. However, with a limited series, 
transient behavior is often encountered. It is clear from Fig. \ref{fig:an-ratio}, that if $x_{max}$ is large enough, the beginning of the series looks like its $x_{max} \rightarrow \infty $ limit which grows at a factorial rate. Consequently, in order to attempt to probe the asymptotic behavior, we will only consider $x_{max}$ such that the available coefficients are 
significantly smaller than their value when $x_{max}$ is infinite. For this reason, we have limited the range of investigation to $x_{max}<1.3$. 
If we have a finite radius of convergence $\lambda_c$, we expect that for $k$ sufficiently large, 
\begin{equation}
|E^{(k)}_0(\sqrt{\omega}x_{max})|\propto (P(\sqrt{\omega}x_{max}))^k
\end{equation}
When this is the case, $\lambda_c(x_{max},\omega)=\omega^3/P(\sqrt{\omega}x_{max})$. In
Fig. \ref{fig:logek}, we have plotted   ln$|E^{(k)}_0(\sqrt{\omega}x_{max})|$ versus $k$. We see a clear linear behavior and 
the slope can be interpretated as ln$P(\sqrt{\omega}x_{max})$. We can then study 
$\lambda_c(x_{max},\omega)$ as a function of $x_{max}$. This is done in 
Fig. \ref{fig:lambdac} where a log-log plot shows a clear linear behavior. 
The slope is near 5.9 which is close to the value 6
that, we will argue in Sec. \ref{sec:rad2}, should hold 
in the limit of small $x_{max}$ (see Eq. (\ref{eq:est})). 
We conclude that in this limit, our numerical data suggests 
\begin{equation}
\lambda_c(x_{max}, \omega=1)\simeq 65 \times x_{max}^{-6}\ . 
\end{equation}

\section{The large $x _{\max }$ limit}
\label{sec:large}

In this section, we first discuss the harmonic energy spectrum with vanishing boundary conditions at $\pm x_{max}$ and then treat the anharmonic interactions with the 
usual perturbative methods. We work in the limit where $x_{max}$ is large. This means that if we consider the $n$-th energy level, $x_{max}$ should be much larger than the largest zero of $H_n(x\sqrt{\omega})$. According to Eq. (6.32.5) of Ref.  \cite{szego}, this 
imposes the restriction $x_{max}>>\sqrt{2n/\omega}$ for large $n$. When all the zeroes of 
$H_n$ are well within $[-x_{max},x_{max}]$, the new boundary conditions require changes 
that are exponentially small. It is clear that for any fixed $x_{max}$, this condition 
will be violated for $n$ sufficiently large.

\subsection{The harmonic case}
We first need to calculate the energy eigenvalues at $\lambda=0$, $E_n^{(0)}(\sqrt{\omega}x_{max}^2$), and their 
corresponding wave functions $\Psi_n^{(0)}(x)$. 
The wave function always depends on $x_{max}$, but this will be kept implicit.
We impose the boundary conditions 
$\Psi_n ^{(0)'}(0)=0$ for $n$ even, $\Psi_n ^{(0)}(0)=0$ for $n$ odd, and 
$\Psi ^{(0)}(x_{max})=0$ in both cases. We will not pay attention to 
the overall normalization until the end of the calculation.
When $x _{\max
}\rightarrow \infty $, we have $E_n^{(0)}(\sqrt{\omega}x_{max})\simeq (n+1/2)$ and we will use
\begin{equation}
\epsilon _n(\sqrt{\omega}x_{max})\equiv E_n^{(0)} (\sqrt{\omega}x_{max})-n-1/2
\end{equation}
as our expansion parameter. Again, $\epsilon_n$ is a dimensionless quantity that 
can only depend on  $\sqrt{\omega}x_{max}$ and we will often keep this dependence implicit. 
We write 
\begin{equation}
\Psi_n ^{(0)}(x)\equiv K_n^{(0)}(x){\rm e}^{-\omega x^2/2}\ .
\end{equation}
This corresponds to a choice $G=-\omega x$
in Eqs. (\ref{eq:basic1}-\ref{eq:basic2}) and consequently
\begin{equation}
\label{eq:keq}
-(1/2)K_n^{(0)''}+\omega x K_n^{(0)'}-n\omega K_n^{(0)}=\epsilon_n \omega K_n^{(0)} \ . 
\end{equation}
We then expand 
\begin{equation}
	K_n^{(0)}=K_n^{(0)(0)}+\epsilon_n K_n^{(0)(1)}+ \dots
\end{equation}
When $x_{max}\rightarrow\infty$, $\epsilon _n \rightarrow 0$, and we want to recover 
the usual harmonic oscillator solution. Consequently, 
we set 
$K_n^{(0)(0)}(x)=H_n(\sqrt{\omega}x)$, the $n$-th Hermite polynomial which is obviously a solution of Eq. (\ref{eq:keq}) at order 
0 in $\epsilon_n$.
At order $\epsilon_n$, we have 
\begin{equation}
-(1/2)K_n^{(0)(1)''}+\omega x K_n^{(0)(1)'}-n\omega K_n^{(0)(1)} =\omega H_n(\sqrt{\omega}x),	
\end{equation}
with the boundary conditions 
\begin{equation}
\label{eq:bc}
K_n^{(0)(1)}(0)=K_n^{(0)(1)'}(0)=0\ .
\end{equation} 

Remarkably, this inhomogeneous equation  can be integrated exactly in two steps. 
First, we set 
\begin{equation}
	K_n^{(0)(1)}(x)=H_n(\sqrt{\omega}x)G_n(x)\ .
	\label{eq:herm}
\end{equation}
This removes the explicitly $n$-dependent term and the equation depend only on $G'$:
\begin{equation}
\label{eq:gp}
\hspace{-30pt}
-(1/2)H_n(\sqrt{\omega}x)G_n^{''}+\left[\omega x H_n(\sqrt{\omega}x)-H_n'(\sqrt{\omega}x)\sqrt{\omega}\right] G_n^{'}=\omega H_n(\sqrt{\omega}x),
\end{equation}
We then write $G'(x)=N(x){\rm e}^{\omega x^2}$ which removes the term 
linear in $x$ and allows us to write the l. h. s. as a total derivative.
The solution is then
\begin{equation}
	G_n(x)=-2\omega \int_{0}^x dy (H_n(\sqrt{\omega}y))^{-2}{\rm e}^{\omega y^2}\int_{0}^y dz {\rm e}^{-\omega z^2}(H_n(\sqrt{\omega}z))^2 \ ,
	\label{eq:g}
\end{equation}
One can check that the 
lower bounds of integration imply the boundary conditions of Eq. (\ref{eq:bc}). 
This is obvious when $n$ is even. When $n$ is odd, the $z$ integral is of order $y^3$ when $y\rightarrow 0$ which overcompensates the $y^{-2}$ of the other factors. 
Note that $G_n(x)$ is independent of $x_{max}$.
The integrand of the $y$-integral has a double pole at the zeroes of $H_n$, however it has no single pole. This can be seen by plugging a Laurent expansion of $G'$ with poles of order 1 and 2 about a zero of $H_n$, and using the relation between the first and the 
second derivative of $H_n$ at this zero provided by the defining equation for Hermite polynomials. 
This forces the coefficient of the simple pole to be zero.
Consequently, when we express $G$ as the integral of $G'$, we can go around the double pole either above or below the real line and 
obtain the same result. In other words, we can regularize the $y$-integral by replacing 
$ (H_n(\sqrt{\omega}y))^{-2}$ by $ (H_n(\sqrt{\omega}y)\pm i \epsilon)^{-2}$.
From these considerations, we see that $G(x)$ develops a simple pole when $x$ approaches 
a zero of $H_n$, which compensate the simple zero of $H_n$. Due to the absence of single 
pole in $G'$, there is no logarithmic singularity in $G$.

$\epsilon _n$ can be calculated by imposing the condition that the 
wave function vanishes at $x_{max}$. This translates into the simple equation:
\begin{equation}
\label{eq:epsn}
\epsilon_n(\sqrt{\omega}x_{max})=-1/G_n(x_{max})\ .
\end{equation}
As $\epsilon_n$ increases, the non-trivial zeroes of the wavefunction move toward the origin. The shift of these zeroes is approximately linear in $\epsilon_n$ for large $x_{max}$. For instance, for $n$ =2, the 
shift of the non-trivial zero $x_{(0)}$ obeys the approximate linear relation $\Delta x_{(0)} \simeq -0.145 \epsilon _2$.

We now discuss in more detail the case of the ground state ($n=0$).
It should be noted that in this case the double integral can be expressed  \cite{cookpri} in terms of generalized hypergeometric series
\begin{eqnarray}
\label{eq:doubleint}
		G_0(x)&=& -\omega x^2\  _{2}F_{2}(1,1;2,3/2;\omega x^2)\\ \nonumber
		            &=&-\frac{\sqrt{\pi}}{2}\sum_{k=1}^{\infty}\frac{(\omega x^2)^k}{k\Gamma (k+1/2)}  \ .
\end{eqnarray}
This series defines a function that converges over the entire complex plane. 
The asymptotic form of $G_0(x_{max})$, for $x_{max}$ large, can be worked out by noticing that 
the integral over $y$ in Eq. (\ref{eq:g}) comes mostly from the region $y\simeq x_{max}$ and consequently, 
we can extend the integral over $z$ to infinity with exponentially small errors.
The integral over $y$ can then be approximately performed by expanding the argument 
of the exponential about $x_{max}$. Using Eq. (\ref{eq:epsn}), we conclude that asymptotically
\begin{equation}
\epsilon_0 (\sqrt{\omega}x_{max})\simeq 2 \sqrt{\frac{\omega}{\pi }}x_{\max }e^{-\omega x_{\max }^{2}}
\label{eq:ep}
\end{equation}
This result coincides exactly with the semi-classical estimate given in Ref. \cite{arbacc}.
The validity of Eq. (\ref{eq:ep}) can be checked numerically.
For instance, for $x_{max}=7$ and $\omega=1$, we get $\epsilon_0=4.14\times 10^{-21}$ 
from Eq. (\ref{eq:ep}) which agrees well with $4.10\times 10^{-21}$
obtained numerically. 

\begin{figure}\begin{center}                                
\includegraphics[width=0.7\textwidth]{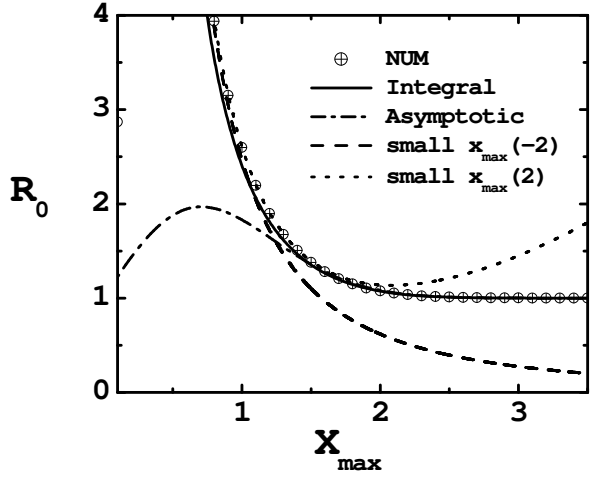}
\caption{Numerical values of $R_0(x_{max})$ (crossed circles) for $\omega=1$, the integral formula of Eq. (\ref{eq:g0}) (solid line), the asymptotic Eq. (\ref{eq:ep}) (dash-dot line), the  leading term ($\propto x_{max}^{-2}$, dash line) of the small $x_{max}$ expansion studied in Sec. \ref{sec:small} and together with the $x_{max}^2$ correction (dots). For $x_{max}>1.5$, the integral formula, its asymptotic form and the data are 
undistinguishable with the eye. \label{fig:r0} }
\end{center}\end{figure}

In Fig. \ref{fig:r0}, we have plotted the asymptotic form Eq. (\ref{eq:ep}) but also the 
full integral form
\begin{equation}
	\epsilon_0 (\sqrt{\omega}x_{max})=1/(
	2\omega \int_{0}^{x_{max}} dy {\rm e}^{\omega y^2}\int_{0}^y dz {\rm e}^{-\omega z^2})\ ,
	\label{eq:g0}
\end{equation}
One notices that the integral form stays accurate to much lower values of $x_{max}$, 
even at very low values of $x_{max}$, where it has no reason to be accurate, it 
gives a reasonable answer. This observation will be discussed further in Sec. \ref{sec:crossemp}.
\subsection{Anharmonic corrections}
Having solved the problem for $\lambda=0$ at first order in $\epsilon _n$, we can use the usual methods of perturbation theory.
$E_0^{(1)}$ can be calculated by 
taking the average of $x^4$ with the corrected ground state wave function constructed above: 
\begin{equation}
\label{eq:e1int}
	E^{(1)}_0=\omega^2N^{-1}\int_0^{x_{max}}dx|\Psi_0^{(0)}(x)|^2x^4\ .
	\end{equation}
with the normalization constant 
\begin{equation}	
N=	 \int_0^{x_{max}}dx|\Psi_0^{(0)}(x)|^2 \ .
\end{equation}
Proceeding as for Eq. (\ref{eq:ep}), we obtain in leading 
order
\begin{equation}
E^{(1)}\simeq (3/4)-\pi^{1/2}\omega^{5/2}x_{max}^5{\rm e}^{-\omega x_{max}^2}
\label{eq:e1}
\end{equation}
Note that in this calculation, we 
have replaced $G_0(x)$ in the integral by its asymptotic form ($\propto x^{-1}{\rm e}^{\omega x^2}$ see Eqs. (\ref{eq:epsn} -\ref{eq:ep})). The 
exponentials exactly cancel inside the integral and we are left with the integration of $x^4/x$. This is justified by the 
fact that most of the contributions come from the large $x$ region. 
The validity of the Eq. (\ref{eq:e1}) can be checked numerically.
For instance, for $x_{max}=7$ and $\omega=1$, Eq. (\ref{eq:e1}) gives a correction $-4.97\times 10^{-18}$ while numerically we obtain $-5.01\times 10^{-18}$.

\begin{figure}
\begin{center}                                
\includegraphics[width=0.7\textwidth]
{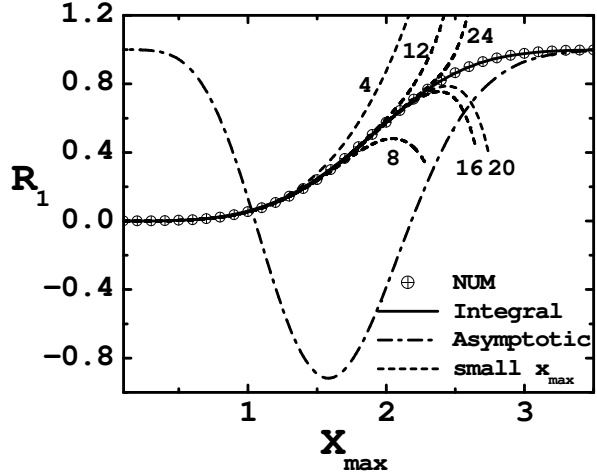}
\caption{
Numerical values of $R_1(x_{max})$ (crossed circles) for $\omega=1$, the integral formula of Eq. (\ref{eq:e1int}) (solid line), the asymptotic Eq. (\ref{eq:e1}) (dash-dot line), the small $x_{max}$ expansion studied in Sec. \ref{sec:small} truncated at order 4, 8, ... 24 (dots). Series analysis suggests that the small $x_{max}$ expansion converges for $x_{max}<2.9$.
\label{fig:r1}}
\end{center}\end{figure}

The good accuracy of Eq. (\ref{eq:e1}) is shown in Fig. \ref{fig:r1}.
In this figure, we also show the full integral formula of Eq. (\ref{eq:e1int}), where the integral was done numerically using 
the square of the order $\epsilon$ approximation  for the 
ground state wave function, without discarding order $\epsilon^2$ terms in the square. Again the integral formula stays valid at unexpectedly low values of $x_{max}$.  This will be explained in Sec. \ref{sec:crossemp}.

\begin{figure}
\begin{center}
\includegraphics[width=0.6\textwidth]{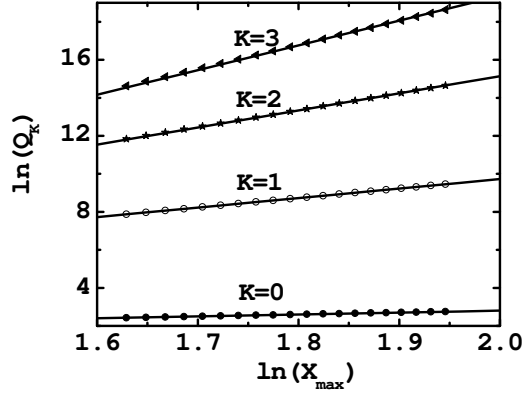}
\caption{${\rm ln}(Q_k(x_{max}))$ versus  ${\rm ln}(x_{max})$ for $\omega=1$. The explanations about the solid lines is given in the text.} \label{fig:pow}
\end{center}
\end{figure}

The derivation of Eq. (\ref{eq:e1}) is quite simple: $\epsilon_0$ give a contribution proportional to $x_{max}e^{-\omega x_{\max }^{2}}$ and the $x$-integral 
gives a factor $x_{max}^4$. 
We conjecture that at leading order, a similar situation is encountered for higher order 
terms, and that
\begin{equation}
\label{eq:conj}
1-R_k(\sqrt{\omega}x_{\max})\propto
x_{\max }^{4k+1}e^{-\omega x_{\max }^{2}}\ .
\end{equation}
For this purpose we have studied numerically the quantity 
\begin{equation}
Q_k({\sqrt{\omega}x_{max}})\equiv e^{+\omega x_{\max }^{2}} (1-R_k(\sqrt{\omega}x_{\max}))\ ,
\end{equation}
which according to the conjecture should scale like $x_{max}^{4k+1}$.
In Fig. \ref{fig:pow}, we have set $\omega =1$ and displayed ${\rm ln}(Q_k(x_{max})$ versus  ${\rm ln}(x_{max})$. The solid lines represent the linear functions 
$A_k+(4k+1){\rm ln}(x_{max})$. The constants $A_k$ have been fitted using the 
last ten data points. Their numerical values are $A_0=0.802$ which compares well 
with the prediction of Eq. (\ref{eq:ep}) ${\rm ln}(4\pi^{-1/2})\simeq 0.814$, 
and $A_1=-0.276$ which compares well 
with the prediction of Eq. (\ref{eq:e1}) ${\rm ln}(4\pi^{-1/2}/3)\simeq -0.285$. 
The linear behavior of the higher orders supports the conjecture.

\section{The small $x _{\max }$ limit}
\label{sec:small}
When the field cutoff $x _{\max }\rightarrow 0 $, the potential term in
the harmonic oscillator is much smaller then the kinetic term, and in first approximation we can use the free particle in
a box of length $2 x _{\max }$. 
The energy levels diverge as $x_{max}^{-2}$ and it is convenient 
to introduce the following rescalings
\[\tilde{H}=x_{max}^2 H\]
\[\tilde{x}=x/x_{max}\]
\[\tilde{\omega}=\omega x_{max}^2\]
\[\tilde{\lambda}=\lambda x_{max}^6\] 
We then obtain a new hamiltonian
\begin{equation}
\tilde{H}=-\frac{1}{2}(\frac{d}{d\tilde{x}})^2+ \tilde{V}\ ,
\end{equation}
with
\begin{equation}
\label{eq:vtilde}
\tilde{V}(\tilde{x})=\left\{
\begin{array}{ccc}
\frac{1}{2}\ \tilde{\omega} ^{2}\tilde{x}^{2}+\tilde{\lambda} \tilde{x}^{4}\quad & {\rm if}& |\tilde{x}| < 1  \\
\infty \quad & {\rm if} & |\tilde{x}| \geq 1
\end{array}
\right.
\end{equation}
With these rescalings, $x_{max}$ has disappeared from the problem and 
we can now expand $\tilde{E}$ as a double series in $\tilde{\lambda}$ 
and $\tilde{\omega}$. This can be done using the usual methods of 
perturbation theory since we can solve the problem exactly when 
$\tilde{\lambda}=\tilde{\omega}=0$. The perturbative series becomes 
\begin{equation}
\tilde{E}_n=\sum_{k=0, l=0}^{\infty}E_n^{(k, l)}\tilde{\lambda} ^k\tilde{\omega}^{2 l}\ ,
\end{equation}
with dimensionless coefficients $E_n^{(k, l)}$. Scaling back to the original problem, we obtain
\begin{equation}
E_n=\sum_{k=0, l=0}^{\infty}E_n^{(k, l)}\lambda ^k\omega^{2 l}x_{max}^{6k+4 l-2}
\end{equation}
Comparing with Eq. (\ref{eq:eexpgen}), we conclude that with this expansion
\begin{equation}
\label{eq:ekn}
E_n^{(k)}(\omega x_{max}^2)=\sum_{ l=0}^{\infty}E_n^{(k, l)}\times (\omega x_{max}^2)^{2 l+3k-1} \ .
\end{equation}
Obviously, this 
implies that the asymptotic behavior when $x_{max}\rightarrow 0$, for the ratios displayed in Fig. \ref{fig:an-ratio}, is 
\begin{equation}
R_k(\sqrt{\omega}x_{max})\propto x_{max}^{6k-2}
\end{equation}
We checked numerically that this asymptotic behavior is correct. Using the data of Fig. \ref{fig:an-ratio}, one can indeed determine the constant of proportionality 
$E_0^{(l,0)}$ calculated independently below with more than two significant digits.
 
\begin{figure}
\begin{center}
\includegraphics[width=0.6\textwidth]{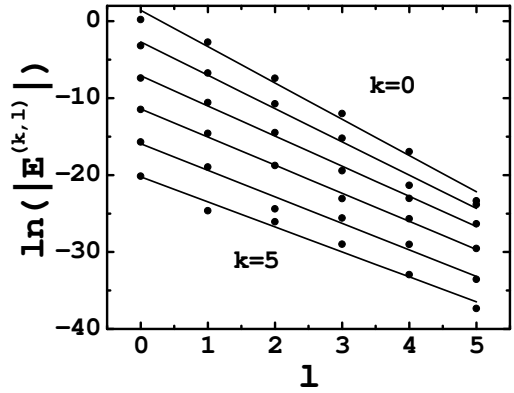}
\caption{$ln(|E^{(k,l)}|)$ versus $l$, the order of expansion in $\omega^2$, for $k=0,\  1,\dots \ 5$. The lines represent linear fits for a given $k$.}
\label{fig:omexp}
\end{center}
\end{figure}

For the ground state, 
\begin{eqnarray}
\label{eq:exactco}
E_0^{(0,0)} &=&\frac{\pi ^{2}}{8} \nonumber \\
E_0^{(1,0)} &=&  \frac{1}{5}  - \frac{4}{{\pi }^2}+ \frac{24}{{\pi }^4}  \nonumber \\
E_0^{(0,1)} &=& \frac{1}{6}-\frac{1}{\pi^2} 
\end{eqnarray}
Higher orders require infinite sums. 
In practice, since we see from Eq. (\ref{eq:vtilde}) that the calculations can be performed with $x_{max}=1$, it is easier to proceed numerically as 
in Sec. \ref{sec:num} but with a double series.
The numerical results are given in Table \ref{table:one} and displayed in Fig. \ref{fig:omexp}.
\begin{table}
\caption{\label{table:co}The coefficients $E_0^{(k, l)}$.}
\label{table:one}
\lineup
\begin{tabular}{@{}c*{5}{c}}
\br
k \verb|\|  l&0&1&2&3\\
\mr
0&  $1.234\m\m\m$ &  $\m6.535 \times 10^{-2}$ & $-5.923\times 10^{-4}$ & $\m6.097 \times 10^{-6}$\\
1&  $\m 4.110 \times 10^{-2}$ & $-1.169\times 10^{-3}$ & $\m2.150\times 10^{-5}$ & $-2.443\times 10^{-7}$&\\
2 & $-6.110\times 10^{-4}$ & $\m2.555\times 10^{-5}$ & $-5.079 \times 10^{-7}$ & $\m3.575 \times 10^{-9}$\\
3 & $\m 1.020 \times 10^{-5}$ & $-4.571\times 10^{-7}$ &$\m6.983 \times 10^{-9}$&$\m9.506 \times 10^{-11}$\\
4 & $-1.517 \times 10^{-7}$ & $\m 5.804 \times 10^{-9}$ & $\m2.492 \times 10^{-11}$ & $-7.641 \times 10^{-12}$ \\
\br
\end{tabular}
\end{table}

\begin{figure}
\begin{center}
\includegraphics[width=0.6\textwidth]{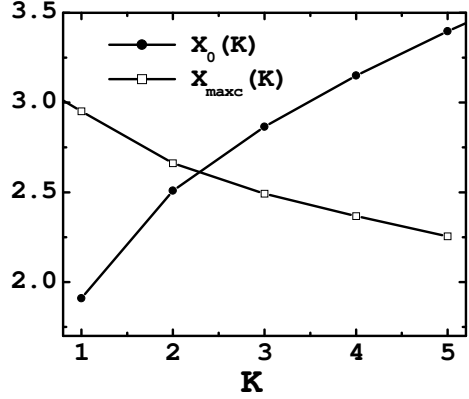}
\caption{Estimated values of $x_{max\  c}(k)$ for which we reach the radius of convergence for $\omega=1$ (empty boxes) and values of $x_0(k)$ (black circles) where the crossover occurs as defined in Sec. \ref{sec:crossemp}, for $k=1,\dots 5$.\label{fig:crossing}}
\end{center}
\end{figure}

The approximately linear behavior of $ln(|E_0^{(k,l)}|)$ in $l$ at fixed $k$ shown in Fig. \ref{fig:omexp} suggests that the power series in $\omega ^2 x_{max}^4$, for the coefficients of $(\lambda/\omega^3)^k$ 
have a finite radius of convergence (see Eq. (\ref{eq:ekn})). As the slope vary from -4.7 for $k=0$ to -3.3 
for $k=5$, we 
expect that the series converge for $|x_{max}\sqrt{\omega}|<{\rm e}^{4.7/4}\simeq 3.3$ for $k=0$ to ${\rm e}^{3.3/4}\simeq 2.3$
for $k=5$. 
Fig. \ref{fig:crossing} shows a steady decrease of the radius of convergence when $k$ increases. At the same time, the value of $x_{max}$ where the transition between the 
small and large cutoff limit occurs, denoted $x_0(k)$ increases with $k$. This quantity 
is defined more precisely in Sec. \ref{sec:crossemp}. Consequently, it seems clear that 
as the order increases, a gap opens between the range of validity of the small $x_{max}$ expansion and the crossover region.
\section{Rigorous lower bounds ont radius of convergence}
\label{sec:rad2}
In Secs. \ref{sec:rad1} and \ref{sec:small}, we have discussed the radius of convergence of series from a numerical point of view. We now return to this question with more 
analytical considerations. First of all, 
it is possible to get an order magnitude estimate 
for the radius of convergence by using a simple argument.
If we look at a typical term of the perturbative series at order $\lambda^k$, it is the 
product of $k$ matrix elements for the perturbation potential $x^4$ divided by $k-1$ 
differences of energy level. The matrix elements of $x^4$ are bounded in magnitude by $x_{max}^4$. 
The differences of energy depend on $x_{max}$. If $x_{max}$ is very small, in other words 
if $x_{max}^2\omega ^2<<\omega$ in units where the mechanical mass and $\hbar$ are set to 1, the harmonic term is a perturbation and the energy differences are proportional to 
$x_{max}^{-2}$. For instance, the 
difference between the ground state and the first excited state will be approximately $3\pi^2/(8x_{max}^2)$. In the opposite limit of 
large $x_{max}$, the level are approximately equidistant with a separation $\omega$. 
From this we expect that  the perturbative series converges for $|\lambda|<\lambda_c(x_{max},\omega)$ with 
\begin{equation}
\label{eq:est}
\lambda_c(x_{max},\omega)\propto \left\{ \begin{array} {ll} x_{max}^{-6}&  {\rm if} \    x_{max} << \omega^{-1/2}\\
\omega x_{max}^{-4} &  {\rm if} \ x_{max} >> \omega^{-1/2}\end{array} \right.
\end{equation}

This argument does not take into account aspects such as the number of terms present or 
their signs, however, a rigorous {\it lower} bound on the radius of convergence
can be obtained by simply using one half of the difference between the ground state and the first excited state of the unperturbed 
system in the above argument. This is formulated  precisely in Theorem XII.11 in Ref. \cite{simonIV} 
and proved in Ref. \cite{kato}. The final result in our case is 
\begin{equation}
\lambda_c(x_{max},\omega)> (1/2)(E_1^{(0)}-E_0^{(0)}  )/x_{max}^4
\label{eq:rig}
\end{equation}
Note that Theorem XII.11 is more than what we need here because the perturbation 
is a bounded operator and we do not need to compare its norm with the norm of $H_0$.

In the case $\omega =0$, dimensional analysis dictates $\lambda_c(x_{max},\omega=0)=C x_{max}^{-6}$ for some positive $C$, and Eq. (\ref{eq:rig}) implies 
\begin{equation}
\lambda_c(x_{max},\omega=0)>(3\pi^2/16)\times x_{max}^{-6}
\end{equation}
This bound can be compared with numerical estimates obtained from the series 
with coefficients $E_0^{(k,0)}$
defined in Sec. \ref{sec:small} and which only requires one calculation with $x_{max}=1$ 
instead of the two step procedure of Sec. \ref{sec:rad1}.
Using the coefficients for $k=5, \dots 10$, we obtain
\begin{equation}
\lambda_c(x_{max},\omega =0 )\simeq 53 \times x_{max}^{-6}\  ,
\end{equation}
which is quite close to result obtained with $\omega =1$ at low $x_{max}$ in Sec. \ref{sec:small}. 
The rigorous bound on $C$, $3\pi^2/16 \simeq 1.85$ is about 30 times smaller than the empirical value 53.  In summary, the lower bound is compatible with our numerical estimate, but is not sharp.

The same argument can be applied for the calculation of $E_0^{(0)}$ as a power series 
in $(1/2)\omega^2$ for the perturbation $(1/2)\omega^2 x^2$. It can be tested by studying the coefficients $E^{(0,l)}$ as we did in Sec. \ref{sec:small}. The rigorous bound becomes  $(1/2)\omega^2_c>(1/2)(3\pi^2/8)/x_{max}^4$, which implies 
\begin{equation}
\omega_c> 1.92\times x_{max}^{-2}
\end{equation}
From Sec. \ref{sec:small}, we have 
\begin{equation}
\omega_c\simeq (3.3)^2 \times x_{max}^{-2} \ ,
\end{equation}
and again, 
the lower bound is compatible with the numerical estimate, but is not an 
accurate estimator of the actual radius of convergence.

\section{The crossover region: empirical data collapse}
\label{sec:crossemp}

In this section, we discuss the question of the interpolation between the large and small 
$x_{max}$ approximation. In Sec. \ref{sec:large}, we have already observed that by naively using integral formulas derived in the large $x_{max}$ approximation 
for $E^{(0)}$ and $E^{(1)}$, it was possible to obtain decent approximations  
at low $x_{max}$. We will first explain how this works and then discuss the higher orders 
terms.

In the (mathematical) limit of very small $x_{max}$, Eq. (\ref{eq:g}) implies that 
$G_0(x)\simeq -\omega x^2$. Consequently, in this limit, where we do not expect the 
approximation to be accurate, the correction becomes dominant and we have  
\begin{equation}
E^{(0)}_0 (\sqrt{\omega}x_{max})\simeq 1/(\omega x_{max}^2)\ .
\end{equation}
Despite the lack of justification for this formula, it is not far from the accurate answer derived in Sec. \ref{sec:small}:
\begin{equation}
E^{(0)}_0(\sqrt{\omega}x_{max})\simeq \pi^2/(8\omega x_{max}^2) \simeq 1.234  /(\omega x_{max}^2)
\end{equation}

Similarly for $E^{(1)}$, we can study Eq. (\ref{eq:e1int}) in the  mathematical limit of small $x_{max}$. According to Sec. \ref{sec:large},
$\Psi^{(0)}_0(x)\simeq (1-(x/x_{max})^2)$ and elementary integration yields
\begin{equation}
E_0^{(1)}(\sqrt{\omega}x_{max})\simeq (1/21)\times(\omega x_{max}^2)^2\simeq 0.0476(\times \omega x_{max}^2)^2
\end{equation}
Again, this is close  to the accurate answer from Eq. (\ref{eq:exactco}):
\begin{equation}
E_0^{(1)}(\sqrt{\omega}x_{max})\simeq 0.0411\times(\omega x_{max}^2)^2
\end{equation}

\begin{figure}
\begin{center}
\includegraphics[width=0.8\textwidth]{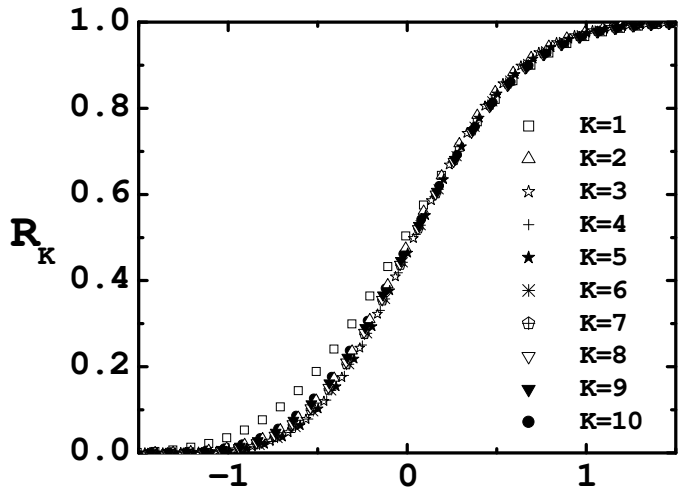}
\caption{$R_k(x_+x_0(k))$ for $k=1, \dots 10$. }
\label{fig:collapse1}
\end{center}
\end{figure}

For higher orders, we know from Fig. \ref{fig:crossing} that the small $x_{max}$ approximation is not convergent in the crossover region. However, 
the shape similarities 
observed in Fig. \ref{fig:an-ratio} suggest to parametrize $R_k$ in terms of a single  
function $U$, that can be shifted by a $k$-dependent quantity that we denote 
$x_0(k)$. We have chosen $x_0(k)$ as the value of $x_{max}$ for which the second derivative of
$R_k$ vanishes. The numerical values are shown in Table \ref{table:xo} and Fig. \ref{fig:crossing}.
\begin{table}
\caption{\label{table:xo} $x_0(k)$ for $k=1\dots 10$.}
\begin{indented}
\lineup
\item[]\begin{tabular}{@{}cc}
\br
& $x_0(k)$\\
\mr
1& 1.908\\
2& 2.509\\
3& 2.868\\
4&3.152\\5& 3.398\\6& 3.626\\7& 3.839\\8& 4.042\\9& 4.235\\10& 4.421\\
\br
\end{tabular}
\end{indented}
\end{table}
Empirically, it can be fitted quite well with 
\begin{equation}
	x_0(k)\simeq 0.87 + 1.13\sqrt{k}\ .
\end{equation}
In Fig. \ref{fig:collapse1}, we show that the possibility of having
\begin{equation}
R_k(x_{max})
	\simeq U(x_{max}-x_0(k))\ ,
\end{equation}
is reasonably well satisfied for $k \geq 2$. In other words, with suitable translations, the $R_k$ approximately ``collapse''.

\section{Conclusions}

We have shown that the calculation of the perturbative series for the anharmonic 
oscillator with a field cutoff $x_{max}$ can be performed reliably using numerical methods, 
and approximate analytical methods in the limit of small and large $x_{max}$. 
For the coefficients of order 0 and 1 in $\lambda$, the integral formulas derived in the large $x_{max}$ limit, produce the correct leading power dependence 
with coefficients close to the correct ones, in the opposite limit (small $x_{max}$) where they are not expected to be accurate. For higher orders, it is possible to 
approximately collapse the crossover region of the various order by an appropriate 
translation in $x_{max}$. The collapse is not perfect and it is necessary to resolve more 
accurately the details of Fig. \ref{fig:collapse1} in order to obtain accurate formulas 
for the $R_k$. 

We expect that it is possible to extend the small and large cutoff techniques in the 
case of higher dimensional field theory, however the only numerical method that can we envision for the crossover region is the Monte Carlo method \cite{continuum03}. 
In view of this, it is essential to reach a proper understanding of the interpolation.

\ack

This research was supported in part by the Department of Energy
under Contract No. FG02-91ER40664 and in part by the National Science Foundation
under Grant No. PHY99-07949. 
This work was completed while Y. M. was visiting the Kavli Institute for Theoretical 
Physics. Y. M. thanks the organizers and participants of the workshop ``Modern Challenges for Lattice Field Theory'' for conversations. We thank J. Cook for checking 
series expansions in Sec. \ref{sec:large} and B. Kessler for providing independent 
checks of the numerical estimates of the radius of convergence.

\section*{References}

\begin{thebibliography}{16}
\expandafter\ifx\csname natexlab\endcsname\relax\def\natexlab#1{#1}\fi
\expandafter\ifx\csname bibnamefont\endcsname\relax
  \def\bibnamefont#1{#1}\fi
\expandafter\ifx\csname bibfnamefont\endcsname\relax
  \def\bibfnamefont#1{#1}\fi
\expandafter\ifx\csname citenamefont\endcsname\relax
  \def\citenamefont#1{#1}\fi
\expandafter\ifx\csname url\endcsname\relax
  \def\url#1{\texttt{#1}}\fi
\expandafter\ifx\csname urlprefix\endcsname\relax\def\urlprefix{URL }\fi
\providecommand{\bibinfo}[2]{#2}
\providecommand{\eprint}[2][]{\url{#2}}

\bibitem
{leguillou90}
\bibinfo{author}{\bibfnamefont{J.~C.} \bibnamefont{LeGuillou}}
  \bibnamefont{and}
  \bibinfo{author}{\bibfnamefont{J.}~\bibnamefont{Zinn-Justin}},
  \emph{\bibinfo{title}{Large-Order Behavior of Perturbation Theory}}
  (\bibinfo{publisher}{North Holland}, \bibinfo{address}{Amsterdam},
  \bibinfo{year}{1990}).

\bibitem
{pernice98}
\bibinfo{author}{\bibfnamefont{S.}~\bibnamefont{Pernice}} \bibnamefont{and}
  \bibinfo{author}{\bibfnamefont{G.}~\bibnamefont{Oleaga}},
  \bibinfo{journal}{Phys. Rev. D} \textbf{\bibinfo{volume}{57}},
  \bibinfo{pages}{1144} (\bibinfo{year}{1998}).

\bibitem
{convpert}
\bibinfo{author}{\bibfnamefont{Y.}~\bibnamefont{Meurice}},
  \bibinfo{journal}{Phys. Rev. Lett.} \textbf{\bibinfo{volume}{88}},
  \bibinfo{pages}{141601} (\bibinfo{year}{2002}{\natexlab{a}}),
  \eprint{hep-th/0103134}.

\bibitem
{bender98}
\bibinfo{author}{\bibfnamefont{C.~M.} \bibnamefont{Bender}} \bibnamefont{and}
  \bibinfo{author}{\bibfnamefont{S.}~\bibnamefont{Boettcher}},
  \bibinfo{journal}{Phys. Rev. Lett.} \textbf{\bibinfo{volume}{80}},
  \bibinfo{pages}{5243} (\bibinfo{year}{1998}).

\bibitem
{gluodyn04}
\bibinfo{author}{\bibfnamefont{L.}~\bibnamefont{Li}} \bibnamefont{and}
  \bibinfo{author}{\bibfnamefont{Y.}~\bibnamefont{Meurice}},
  \bibinfo{journal}{Phys. Rev. D} \textbf{\bibinfo{volume}{71}},
  \bibinfo{pages}{016008} (\bibinfo{year}{2005}{\natexlab{a}}),
  \eprint{hep-lat/0410029}.

\bibitem
  {optim03}
\bibinfo{author}{\bibfnamefont{B.}~\bibnamefont{Kessler}},
  \bibinfo{author}{\bibfnamefont{L.}~\bibnamefont{Li}}, \bibnamefont{and}
  \bibinfo{author}{\bibfnamefont{Y.}~\bibnamefont{Meurice}},
  \bibinfo{journal}{Phys. Rev.} \textbf{\bibinfo{volume}{D69}},
  \bibinfo{pages}{045014} (\bibinfo{year}{2004}), \eprint{hep-th/0309022}.

\bibitem
{plaquette}
\bibinfo{author}{\bibfnamefont{L.}~\bibnamefont{Li}} \bibnamefont{and}
  \bibinfo{author}{\bibfnamefont{Y.}~\bibnamefont{Meurice}},
  \bibinfo{journal}{Phys. Rev.} \textbf{\bibinfo{volume}{D71}},
  \bibinfo{pages}{054509} (\bibinfo{year}{2005}{\natexlab{b}}),
  \eprint{hep-lat/0501023}.

\bibitem
  {buckley92}
\bibinfo{author}{\bibfnamefont{I.~R.~C.} \bibnamefont{Buckley}},
  \bibinfo{author}{\bibfnamefont{A.}~\bibnamefont{Duncan}}, \bibnamefont{and}
  \bibinfo{author}{\bibfnamefont{H.~F.} \bibnamefont{Jones}},
  \bibinfo{journal}{Phys. Rev.} \textbf{\bibinfo{volume}{D47}},
  \bibinfo{pages}{2554} (\bibinfo{year}{1993}).

\bibitem
{duncan92}
\bibinfo{author}{\bibfnamefont{A.}~\bibnamefont{Duncan}} \bibnamefont{and}
  \bibinfo{author}{\bibfnamefont{H.~F.} \bibnamefont{Jones}},
  \bibinfo{journal}{Phys. Rev.} \textbf{\bibinfo{volume}{D47}},
  \bibinfo{pages}{2560} (\bibinfo{year}{1993}).

\bibitem
{bender69}
\bibinfo{author}{\bibfnamefont{C.}~\bibnamefont{Bender}} \bibnamefont{and}
  \bibinfo{author}{\bibfnamefont{T.~T.} \bibnamefont{Wu}},
  \bibinfo{journal}{Phys. Rev.} \textbf{\bibinfo{volume}{184}},
  \bibinfo{pages}{1231} (\bibinfo{year}{1969}).

\bibitem
{arbacc}
\bibinfo{author}{\bibfnamefont{Y.}~\bibnamefont{Meurice}}, \bibinfo{journal}{J.
  Phys.} \textbf{\bibinfo{volume}{A35}}, \bibinfo{pages}{8831}
  (\bibinfo{year}{2002}{\natexlab{b}}), \eprint{quant-ph/0202047}.

\bibitem
{szego}
\bibinfo{author}{\bibfnamefont{G.}~\bibnamefont{Szego}},
  \emph{\bibinfo{title}{Orthogonal Polynomials}} (\bibinfo{publisher}{American
  Mathematical Society}, \bibinfo{address}{Providence}, \bibinfo{year}{1939}).

\bibitem
{cookpri}
\bibinfo{author}{\bibfnamefont{J.}~\bibnamefont{Cook}}, \bibinfo{note}{private
  communication}.

\bibitem
{simonIV}
\bibinfo{author}{\bibfnamefont{M.}~\bibnamefont{Reed}} \bibnamefont{and}
  \bibinfo{author}{\bibfnamefont{B.}~\bibnamefont{Simon}},
  \emph{\bibinfo{title}{Methods of Modern Mathematical Physics: IV Analysis of
  Operators}} (\bibinfo{publisher}{Academic Press}, \bibinfo{address}{San
  Diego}, \bibinfo{year}{1978}).

\bibitem
{kato}
\bibinfo{author}{\bibfnamefont{T.}~\bibnamefont{Kato}},
  \emph{\bibinfo{title}{Perturbation Theory for Linear Operators}}
  (\bibinfo{publisher}{Springer Verlag}, \bibinfo{address}{New York},
  \bibinfo{year}{1966}).

\bibitem
{continuum03}
\bibinfo{author}{\bibfnamefont{L.}~\bibnamefont{Li}} \bibnamefont{and}
  \bibinfo{author}{\bibfnamefont{Y.}~\bibnamefont{Meurice}},
  \bibinfo{journal}{Nucl. Phys. Proc. Suppl.} \textbf{\bibinfo{volume}{129}},
  \bibinfo{pages}{883} (\bibinfo{year}{2004}), \eprint{hep-lat/0309066}.

\end{thebibliography}

\end{document}